# The Ethics of AI in Education

Chapter 26

Handbook of Artificial Intelligence in Education

Edited by Benedict du Boulay, Antonija Mitrovic, Kalina Yacef

Kaśka Porayska-Pomsta
UCL Knowledge Lab,
UCL'S Faculty of Education and Society
University College London
UK
ORCID: 0000-0002-9433-4022

Wayne Holmes
UCL Knowledge Lab,
UCL'S Faculty of Education and Society
University College London
UK
ORCID: 0000-0002-8352-1594

Selena Nemorin
Department of Education
Kellogg's College
University of Oxford
UK
ORCID: 0000-0002-8668-8859

**Abstract.** The transition of Artificial Intelligence (AI) from a lab-based science to live human contexts brings into sharp focus many historic, socio-cultural biases, inequalities, and moral dilemmas. Many questions that have been raised regarding the broader ethics of AI are also relevant for AI in Education (AIED). AIED raises further specific challenges related to the impact of its technologies on users, how such technologies might be used to reinforce or alter the way that we learn and teach, and what we, as a society and individuals, value as outcomes of education. This chapter discusses key ethical dimensions of AI and contextualises them within AIED design and engineering practices to draw connections between the AIED systems we build, the questions about human learning and development we ask, the ethics of the pedagogies we use, and the considerations of values that we promote in and through AIED within a wider socio-technical system.

## 1. Introduction

The advent of big data, and of Artificial Intelligence (AI) applications that collect and consume such data, has led to fundamental questions about the ethics of AI designs and to efforts aimed to highlight and safeguard against any potential harms caused by the deployment of AI across diverse domains of applications. Typically, questions raised relate to the *trustworthiness* of AI as agent technologies that autonomously or semi-autonomously operate in human environments and that have the ability to alter human behaviour. Other questions concern the role that AI may play now and in the future in either resolving or amplifying pre-existing social biases and any resulting harms. Specifically, *Ethical AI* as an emergent area of AI research and policy, has been spurred by the revelations of AI applications (usually unintentionally) promoting and amplifying many of the discriminatory and oppressive practices, and assumptions that underpin pre-existing social and institutional systems, e.g., historical biases against non-dominant populations, against users characterised by some divergence from the so-called cognitive or physical 'norm', or those who are socio-economically disadvantaged (Crawford, 2017a; Madaio et al., 2022; Porayska-Pomsta and Rajendran, 2019; Williamson, Eynon, Knox & Davis, *in this volume*). Numerous examples of AI bias are both well-documented and rehearsed throughout the emergent ethics of AI literature, in hundreds of policy reports about AI ethics and governance that have been published to date (c.f. Jobin, Ienca & Vayena 2019; Hargendorff, 2020) and in the media. Despite this, our understanding of the ethics for AI in Education (AIED) is still fledgling (Holmes *et al.*, 2021).

One reason why the ethics of AIED have received little attention to date may be because the field has traditionally adopted a self-image of being inherently 'good' in its intentions and as such it has implicitly assumed that these good intentions are automatically encoded in the AIED technologies (Holmes *et al.*, 2021). This positive self-image stems from two further face-value assumptions tacitly adopted by AIED researchers, namely that: (a) AIED is by default simultaneously guided by and guarded against any potential pitfalls and unintended consequences by the intentions and practices of the broader education system; and (b) AI technology can serve to promote social justice, inclusive good quality education for all. However, neither assumption has so far received much critical attention from the community and few attempts have been made to examine the ethical value, safety, and trustworthiness of AIED systems, approaches, and methods in the broader socio-technical context.

The first assumption reflects AIED researchers' historic surrender of the responsibility for reflecting on and addressing the questions about the ethical implications of their AIED intervention designs and use to the decision-makers within the broader education system which they intend to fit and serve. The second assumption derives from a particular approach in formal education, which proved convenient for the early computer assisted learning technologies, namely the drill and practice mastery learning focused pedagogies. This approach is deeply ingrained in traditional educational practices focused on school-based subject-domain teaching, which was further boosted by Bloom's influential "2-sigma effect" studies (Bloom, 1984). These studies have served as the foundation for much AIED work, with the mastery learning model of education and the perceived benefits of one-to-one individualised, adaptive teaching support providing the ultimate ambition for the AIED technologies to strive for, and against which to measure their success. However, as in other domains of AI applications such as healthcare, just because AIED explicitly aims to fit into a pre-existent system does not *de facto* guarantee that the practices that it promotes are ethical. Similar to other domains, both the possible systemic biases and the domain-dependent definitional idiosyncrasies related to concepts such as fairness, equity, or human autonomy need to be determined and examined in the context of AI applications for Education, to allow the AIED engineers and diverse users of AIED technologies to develop an understanding of the ethics of AIED systems' designs, and to guide those designs and their deployments accordingly.

In this chapter we discuss concepts that have emerged in the broader AI applications' contexts as key to the ethics of AI, and we contextualise them in the discussions of and directions taken in the ethics of AIED research. First, we consider briefly what might count as an 'ethical approach', in order to ground and examine these concepts specifically within AI and AIED. We recognise that AIED provides a unique domain in which to study the ethics of AI, not least because of its central focus on supporting the interaction with human cognition and on delivering pedagogies that nudge learners towards long-term learning behaviour changes. As such the examination of the assumptions, approaches and methods employed within the discipline of AIED has the potential to lead to a greater understanding of whether and how AIED systems are ethical, of the blind spots and areas for improvement for the field with respect to ethics, and to insights from the AIED of potential importance to other areas of AI that intend to influence or enhance human cognition, decision-making, and behaviour. We review key Ethical AI concepts, which provide the foundations for the study of Ethical AIED. Next, the broad Ethical AI concepts are contextualised within AIED, and the unique aspects of the AIED domain are highlighted. The penultimate section, provides an outline of an Ethics of AIED framework and offers detailed initial mapping between different forms of bias in AI, AIED and broader socio-technical context (Table 2). We conclude the chapter with a brief examination of the gaps and the future directions for Ethical AIED.

## 2. What counts as an ethical approach?

AIED's self-image of being inherently good, for the assumptions outlined above, begs the question: what does 'being good' actually mean? Are there immutable, universal ethical principles that should guide individual and collective behaviour; or does what constitutes being good depend on one's individual socio-political perspective? Such questions are inevitably complex and remain open despite more than two thousand years of ethics discourse. The emergent research on Ethical AI provides a starting point for considering the necessary, even if not sufficient, principles of Ethical AIED. Specifically, within the broader context of AI, a consensus has emerged that any attempt to develop robust and actionable principles that ensure AI research and practices are 'good' ought to be grounded in core concepts from moral philosophy, with special emphasis on universal human rights and obligations (Kant, 1785; Ross, 1930). Such obligations have been recently re-conceptualised specifically in the context of AI by Floridi and Cowls (2019) as: (i) **beneficence** ('do good' for human wellbeing, dignity and for the planet); (ii) **non-maleficence** ('do no harm' by avoiding over-reliance, over-use, or misuse of AI technologies, in order to preserve personal privacy of users, and to prevent use for harmful purposes); (iii) **autonomy** (promote human autonomy, do not impair human freedom of choice and decision-making, by delegating decisions to AI); (iv) **justice** (seek and preserve justice, prevent any forms of discrimination, foster diversity, including the way in which AI is used to enhance human decision-making); and (v) **explicability**, i.e. *transparency* with respect to how AI works, and *accountability* relating to who is responsible for how AI works. The explicability principle is considered an enabler for applying the first four principles, insofar as it gives access to how a given AI technology allows the user to exercise their autonomy and to be audited with respect to any potential benefits and harms. All five principles are considered necessary to support trustworthy and responsible AI designs and deployment. Floridi and Cowls' principles aim for universality, domain-independence, and to serve as a basis for a dialogue between engineering and social scientific perspectives on AI. However, this comes at the price of a lack of concreteness. In short, there remains a paucity of guidelines for how those principles might be actioned in specific AI designs and application contexts. A substantial part of the problem lies in the relativistic nature of the concepts involved, which tend to depend on: (a) socio-cultural norms, which are in themselves subject to change over time (e.g. the concept of justice is perpetually evolving), (b) circumstances and needs of individuals (e.g. the concept of individual fairness is deeply rooted in the situational and subjective realities of individuals), and (c) context (e.g. the tension between multiple conflicting interests, expectations and needs of different stakeholders, which may lead to different interpretations of the beneficent nature of AI).

## 3. Ethics of AI: key dimensions and concerns

One way in which researchers are trying to understand how the overarching ethics principles might be considered in AI systems' designs and their deployment is by exploring the exact sources and the potential consequences of principles of ethical AI being violated. This allows for the specific related harms to be considered and articulated, and to highlight the complex relationships between data that encode historical inequalities and the socio-cultural biases in the decisions affecting different groups. Although the related work so far has improved our understanding of the different forms of societal and systemic biases and of how such biases may be both reinforced through and mitigated within AI algorithms, researchers in this area are still faced with substantial challenges arising from different disciplinary perspectives involved (e.g., moral philosophy vs. law, vs. social justice, vs. AI engineering, and so on), each perspective bringing different definitions of the key terms (e.g., sociological vs. statistical understanding of bias). Bias in AI has long been the subject of research, revealing overlapping definitions and different points at which it may enter the AI development and deployment pipeline (Crawford, 2017a; Blodgett, Barocas, Daumé & Wallach 2020; Baker, Ocumpaugh & Andres, 2021). In this context, it remains critical for AI practitioners to question when and how bias may enter into AI systems that they build, and how to mitigate or eliminate it.

### 3.1 General forms of bias

Bias has emerged as a concept that underpins key ethical concerns in AI. There are many different conceptions of bias available through diverse science disciplines and schools of thought. Fundamentally, bias usually refers to people's tendency to stereotype situations, groups, or individuals (Cardwell, 1999), and this ultimately leads to people's tendency to favour certain things or people over others. One common feature of the different definitions of bias is their reference to some form of *discrimination* that creates a disparity in the treatment of different groups of people or individuals. Danks & London (2017) distinguish between: (a) **moral bias**, i.e., a deviation from certain moral principles related to equity, autonomy and human rights; (b) **legal bias**, i.e., undue prejudice, such as judgments based on preconceived notions of particular groups, which violates written legal norms prohibiting discrimination, and (c) **statistical bias**, i.e., a flaw in the data, in the data collection process, or in experimental design, which generates results that do not represent accurately the population at large, either because of *under-* or *over-* representation. These different types of bias do not always align and thus, addressing one type of bias may not render a system ethically 'better'. For example, it is possible for data collection to be statistically unbiased and to generalise well to new data. Nevertheless, if the data collection takes place in the context where there is historical bias, this training data could lead to biased decisions in the legal and moral sense, due to structural inequalities encoded into these data. An algorithm might also not be statistically or legally biased, but it might be morally deviant insofar as it (re)produces structural inequalities or reinforces stereotypes that violate common ideals of equity, e.g., use of gendered language models that accurately reflect how language is used in society, but that also reinforce historically ingrained gender stereotypes.

## 3.2 Algorithmic bias and sources of bias in AI systems

Much has been said to date about bias that is encoded in data and that is inherited from people's attitudes and beliefs (Crawford, 2017b). Recently, with the greater use of AI systems in diverse mainstream application contexts, increased attention has been dedicated to the AI systems' development process during which different forms of bias can be introduced. In this broader context, bias is referred to as *algorithmic bias*.

Algorithmic bias is predominantly studied in the context of the AI subdomain of machine learning, with the machine learning system development process, from task definition to deployment of an AI system, being referred to as the *machine learning pipeline*. Given that machine learning is a statistical inference method, algorithmic bias is typically understood as statistical bias. In this subsection we synthesise the key considerations with respect to algorithmic bias, as studied within the context of machine learning, looking specifically at the sources of bias and the steps in the AI systems' development at which bias may be introduced. We also provide some specific recommendations made in the literature for best practices aimed to mitigate or reduce bias in AI systems.  However, many considerations presented here also apply to AI rule/knowledge-based approaches, including to the knowledge representation models and the types of heuristics that are used to drive inference in such systems.

Cramer, Vaughan & Holstein (2019) propose a seven-step machine learning pipeline, involving (a) **task definition**, concerned with identifying and specifying the problem an AI system is designed to address, e.g. to help predict student attainment; (b) **data construction**, which involves selecting a data source, acquiring data, pre-processing and labelling data; (c) **model definition**, which involves the selection of a specific AI approach and of the objective function; (d) **training process**, when the model is trained on data; (v) **evaluation process**, when the model is validated on additional data and where there is an opportunity to check for biases in the entire system; (e) **deployment**, when the system leaves the lab and where any mismatches between training data and target populations become apparent, and (f) **feedback** on how the system fares in the wild, based on the way that it is being used. Bias can enter, or it can be reinforced, at any step in this pipeline, with decisions made during earlier steps also affecting the ethical quality of the steps downstream. For example, the choice of a task (part of step 1) may affect whether an entire system is ethical, as demonstrated by the infamous case of the Wu and Zhang (2016) classification system designed to predict people's criminality based on their facial features.

*Table 1 Main types of AI bias, their potential sources, and related harms*

| AI Bias | | |
|---|---|---|
| **Type** | **Sources** | **Harms** |
| **Historic** *(Diffused through time, cultures and societies)* | Moral, legal and socio-cultural biases | Representational (especially recognition, denigration and exnomination); allocative, outcome; process; individual; group fairness; human autonomy |
| **Representation** *(Over-/under-representation of some populations; methodological biases)* | Moral, legal, historic; assumptions and methods biases | Representational; allocative, outcome; process; individual; group fairness |
| **Measurement** *(Choices of features and labels and how these should be computed)* | Variation between measurement methods, including accuracy of measurement, used for different groups | Representational; allocative; individual; group fairness |
| **Aggregation** *(Assumption of mapping consistency between inputs and labels)* | Model's inability to account for data other than the data that it was trained on; untested assumptions of mapping consistency | Representational (especially recognition and exnomination); outcome; individual and group fairness |
| **Learning** *(Performance differences resulting from different choices of modelling and model evaluation/objective functions)* | Assumptions and methods biases; mal-informed/unverified goal satisfaction priorities | Representational; outcome; process; individual; group fairness |
| **Evaluation** *(Failure of the benchmark data to match the use population data; lack of generalisablity; overclaimed quality of models)* | Historic, context; representation; benchmark data bias | Representational; allocative; outcome; process; individual; group fairness |
| **Deployment** *(Mismatch between the task the model is designed to solve and the task for which it is being used)* | Assumptions bias; mal-informed/unverified definition of deployment context; techno-centric methods and focus | Representational; allocative; outcome; process; individual; group fairness |

Nonetheless, data are considered the main source of algorithmic bias. Data bias may be introduced: (a) ***at the source***, e.g., during sampling, (b) ***during pre-processing of data***, e.g., during data cleaning or labelling, or (c) *through the **data collection method***, including the specific data collection software, sampling strategy, or human interpretation of responses (e.g., when qualitative methods such as semi-structured interviews are used). At each point in the process there is a need for AI researchers and designers to assess their decisions for their ethical value, by carefully considering the provenance of the data, how the data has been acquired, and whether the handling of the data (pre-processing and labelling) is trustworthy and ethically sound (Cramer et al., 2019). Researchers have also identified different forms of data bias. For example, Suresh and Guttag (2020) discuss seven categories (listed in Table 1 along with their potential sources and harmful consequences):

**Historic bias**, which typically occurs at the point at which data is being generated, refers to bias that is deeply embedded in historic, cultural, and social stereotypes that discriminate against particular groups (e.g., word embeddings and gendering of nouns describing professions such as female nurse vs. male doctor).

**Representational bias** refers to lack generalisability of the sample data to different populations due to certain populations being under-represented, over-represented, or due to sampling methods that are uneven or limited. For example, this form of bias can be found in systems such as ImageNet (a dataset of labelled images), which are based on data collected predominantly in specific socio-geographical settings (e.g., North America), with limited data from the rest of the world.

**Measurement bias** typically occurs during decisions related to what features and labels should be used in a model and to how such features should be collected and computed. Since labels are proxies for often complex and abstract constructs (e.g., creditworthiness or student engagement), they are inevitably oversimplifications of the real things. Furthermore, measurement bias may occur when the measurement method, or when the accuracy of measurement, vary between groups, e.g., when more stringent assessment of students is applied in one school district compared to another.

**Aggregation bias** happens when a model cannot account for the data that should be considered differently from the data on which the model was trained (e.g., when a model derived from data of neurotypical learners is used for neuro-divergent students). This type of bias results from an assumption that the mappings between inputs and the labels are consistent across different subsets of data. A model harbouring aggregation bias may either only fit a dominant population, or it may not be optimal for any group.

**Learning bias** is introduced when modelling choices, such as the choice of an objective function in machine learning, leads to performance differences across different data. An objective function encodes the goals of an agent and a measure of accuracy of a model (according to assumptions that may in themselves be biased), so prioritising one objective over another may lead to inaccurate classification and even dangerous outcomes. For example, the infamous case of the Microsoft Tay twitter chatbot illustrates how the objective of increasing user engagement at all cost may lead to system's morally deviant behaviours such as racist abuse.

**Evaluation bias** occurs when the so-called benchmark data fails to match the use population data. This is a known issue with respect to models that are optimised for training data, but which fail to generalise to new data. The issue is exacerbated by the fact that models are frequently evaluated against each other, which can lead to unsubstantiated generalisations about how good the specific models actually are in real world contexts. A key question here relates to whether the benchmark data harbours historical, representation, or benchmark biases that may be obscured by lack of evaluations against real world scenarios.

**Deployment bias** refers to a mismatch between the task that the model is supposed to solve and the task for which it is being used. This form of bias can be hard to control, since users can and often do appropriate technologies in ways that may not have been intended by their designers. This may introduce bias and unintended harms, for example when users interfere with or override system decisions, or when a system is presented as fully autonomous when in fact it requires human intervention to reach its goals. Risk assessment tools used in the criminal justice context offer examples of this form of bias, providing a warning of potential similar dangers in the context of learning analytics.

Cramer et al., (2019) offer important recommendations for what AI designers need to consider in order to address different sources of data bias. For example, with respect to examining the data source, it is important to consider the possibility of any potential societal or cultural basis for data bias such as a tendency to use gendered language to describe professions (again, female nurses vs. male doctors), and to consider carefully whether and how exactly data sources match the system's intended deployment contexts. Regarding bias that results from pre-processing and labelling of data, Cramer et al. recommend examining data for any biases that may be the result of *discarding data* (e.g., when someone may not want to declare their gender or ethnicity), *bucketing values* (e.g., if someone identifies with more than one race), *using pre-processing and labelling software* which may already harbour bias (e.g., google translate assigning gender to profession nouns), and *human labellers* who make inherently subjective judgements.

With respect to the data collection methods, it is critical to audit carefully the pre-processing tools for bias and to develop and employ techniques to quantify and reduce bias introduced by human labellers. It is also key to ensure that data is representative of target users, and to be transparent about any known representational limitations in data. Finally, it is also important to question whether the data collection process is ethical in itself. For example, efforts to limit representational bias may also lead to overburdening under-represented groups. Considering what methods (if any) may help reduce or eliminate such a 'participation tax' should be integral to any ethical data collection practices.

## 3.3 Harm

Bias is considered the root cause of diverse forms of *harms*. The different forms of bias have been linked to specific forms of harmful consequences for those individuals and groups against whom the bias is tilted. Barocas, Crawford, Shapiro & Wallach (2017) identify two overarching categories of harms: (a) harms of allocation, i.e., harms that relate to certain opportunities or resources being withheld from some groups or individuals, and (b) harms of representation, i.e., usually stereotypical and negative ways, in which certain groups may be represented or in which they may not be represented in a positive light.

***Harms of allocation*** have been studied in greater depth than harms of representation, since they are also often considered from an economic perspective. Examples of allocative bias and harms include restrictions imposed on young people below a certain age on getting mortgages or high-achieving students in schools in poor socio-economic communities being graded as lower achieving, based on a grade average for the schools[1]. Allocative harms are transactional in nature. They are immediate, easily quantifiable, and time-bound. Thus, harms of allocation raise questions of fairness and justice that can be examined with respect to precise and discrete transactions or decision incidents.

By contrast ***harms of representation*** refer to a relatively neglected area of study within computer science and AI, mainly because such harms are not discrete and they do not occur in single transactions, instead being ingrained within cultural and institutional contexts. As such representational harms are long-term processes that impact people's beliefs and attitudes. They are diffused across time, diverse historical and socio-cultural contexts. They often lie at the root of allocative harms (Crawford, 2017b).

Barocas *et al.* (2017) specify five sub-types of harms of representation that need to be considered in the context of algorithmic bias: (a) ***stereotyping***, e.g., gender stereotyping used in translation algorithms (Boloukbasi, Chang, Zou, Saligrama & Kalai, 2016); (b) ***recognition***, which involves certain groups being erased or made invisible to the algorithm (Sweeney, 2013), such as AI technologies not being able to recognise users of colour (Buolamwini & Gebru, 2018); (c) ***denigration***, which involves users' dignity being violated by an algorithmic bias, with the infamous example of black people not being recognised as humans by the Google picture classification algorithm (Alciné, 2015); (d) ***under-representation***, which refers to certain groups being systematically ignored as representative of, for example, certain professions such as women as CEOs; (e) *ex-nomination*, where the majority demographic becomes accepted as a norm thereby amplifying any difference as an undesirable deviation, e.g., neurodiversity in a mainstream education system that is geared towards neuro-typicality.

---

[1] https://www.theguardian.com/education/2020/aug/13/england-a-level-downgrades-hit-pupils-from-disadvantaged-areas-hardest

The two overarching types of harms (of allocation and representation) and the subtypes of representational harms are not only interrelated, but also multiple types of harms may be embedded in any given system design, in the way that it is deployed, and in the impact that it might have on different users. Such impact, and bias more broadly, is inextricably linked to questions of *fairness* and *justice* and of how technology contributes to mitigating or exacerbating social inequalities.

### 3.4 Fairness

The questions about algorithmic bias cannot be separated from the questions about what constitutes *fairness* (Kizilcec & Lee, 2022; Narayanan, 2018). Fairness in the context of technology concerns what and whose values are embedded into the technology, who is excluded as a result, and whether this is justified in the eyes of the law, in the light of moral principles, and in the eyes of the individuals who may be affected by the way in which a given technology operates. The notion of fairness has been studied within moral and political philosophy for centuries (see e.g., Rawls' (1958) justice as fairness). Recently, the notion of fairness has also become a subject of particular interest in AI research and engineering. In the AI engineering context, one aim is to furnish algorithms with *fairness metrics*, through quantifying the notion of fairness, to enable appropriate mitigation of any biases and related harms. In this context, there is an inevitable tension between the social and computational sciences conceptions of fairness. Attempting to mathematically quantify a context- and perspective-dependent notion of fairness is considered non-trivial, if not doomed to failure. To paraphrase Narayanan, fairness cannot be equated with the number 0.78 (or any other number for that matter), however hard computer scientists might try, because of the definitional multiplicity of the construct of fairness, which is socio-culturally, institutionally, situationally, and subjectively determined (Barocas *et al.*, 2021; Narayanan, 2018).

To illustrate his point, Narayanan identifies at least 21 different definitions of fairness within computer science alone, depending on whose perspective one assumes: (a) the perspective of the designer, in which the fairness-related questions asked of an algorithm will concern its *predictive value*, i.e., the algorithm's ability to classify correctly data with respect to some feature of interest, (b) the perspective of the person affected and who may or may not agree with a given decision , or (c) the perspective of society, in which the question of fairness is considered in terms of its benefits for a wider group and where the interests of any given individual may be overridden. With respect to society's perspective, any evaluation of what is and is not fair will be made based on what is considered socio-culturally acceptable within any given socio-historical context. In particular, if specific biases form part of a given culture or society's habitual perceptions (e.g., the perception of inferiority of some social groups as in racial discrimination, or deviation from some culturally established notion of norm as is routinely experienced by neurodivergent groups) then these biases will inevitably infiltrate any local interpretation of fairness and the related actions. Thus, fairness is by no means a neutral notion. Hence the questions we ask about fairness and the way in which we approach algorithmic fairness cannot simply be about mathematical neutrality, but rather about people's preferences, judgements, beliefs, desires, and needs – with an accompanying explicit acknowledgement that, when it comes to different perspectives, neutrality is virtually impossible.

The challenges of furnishing AI systems with fairness are highlighted when we consider the different types of fairness at play. Discussions about AI fairness centre on two dimensions: (a) ***individual fairness*** (judged in terms of the outcomes for individuals) vs. ***group fairness*** (judged in terms of outcomes for groups of individuals defined along some common identity criteria), and (b) ***outcomes fairness*** (i.e., *equality of the results* of certain processes for groups or individuals) vs. ***process fairness*** (i.e., *equality of treatment* defined by the factors that bear on how the specific processes come about and are undertaken). Friedler, Scheidegger & Venkatasubramanian (2016) demonstrate that reconciling these dimensions is effectively impossible, since the different perspectives often lead to conflicting interpretations of what is and what is not fair: what may seem fair for one group, may be deemed unfair for another group, or it may seem completely unfair to specific individuals. Hence, there is an acute need in AI engineering to make explicit the diverse and often conflicting assumptions that underpin the particular conceptions of fairness embedded in the system. The nature of the goals of any given system, its efficacy in achieving those goals, the target users, the context for which a system is being designed and in which it is being deployed and evaluated, all need to be made *transparent*. Simultaneously the system's operation needs to be made *explainable* for the users, for the systems to be rendered trustworthy and their designers and decision-makers to be made accountable for the outcomes resulting from their systems' use.

## 4. Ethical considerations for AI in Education

The key ethical considerations reviewed, including the specific issues related to algorithmic bias, the associated harms, and fairness, are of direct pertinence to AIED (see also Williamson et al., *in this volume*; Brooks, Kovanovic & Nguyen, *in this volume*). As an academic field, AIED inherently aims to contribute to fairer, more equitable, more accountable, and more educated global society either by helping to reduce attainment gaps (O'Shea, 1979; Reich & Ito, 2017; VanLehn, 2011), or by addressing specific gaps in existing education systems (Saxena, Pillai & Mostow, 2018; Uchiduino et al., 2018; Madaio et al., 2020; Holstein and Doroudi, 2022). Driven by those aspirations and shaped by its central focus on supporting and enhancing human learning, the field has invested in design, usage, and evaluation practices (some, decades ahead of other AI subfields), that are explicitly aimed to generate evidence of the efficacy and safety of the AIED systems for human learners, e.g., through methods adopted from the psychological and learning sciences. As a design discipline, the field has also invested in increasing the relevance of the AIED technologies to diverse users, e.g., through the application of user-centred and participatory design methods (Porayska-Pomsta et al., 2013; Porayska-Pomsta & Rajendran, 2018), and in improving the accuracy, transparency and explainability of the underlying models, notably, through *glass box* approaches such as the open learner models (Bull and Kay, 2016; Conati, Porayska-Pomsta & Mavrikis, 2018; see also Kay, Kummerfeld, Conati, Holstein, & Porayska-Pomsta, *in this volume*). AIED has therefore an important contribution to offer to the wider AI community with respect to the methods, practices and examples of rigorous research that respond to many Ethical AI research recommendations and concerns (Conati *et al.*, 2018). However, the *a priori* intended beneficence of the AIED has also led to a certain level of complacency with respect to identifying and addressing any real and potential ethical blind spots for the field (Holmes *et al.*, 2021), which are now increasingly being identified and examined by the AIED community (Holmes & Porayska-Pomsta, 2022).

In their overview of the key issues surrounding the ethics of AI for education, Holstein and Doroudi (2022) identify two overarching levels at which questions about the ethics of AIED need to be addressed: (a) the **socio-technical level,** including the questions about equity of access to and benefits from using technology, and about the mutual influence that the socio-technical system and AI exert on each other; and (b) the **AIED Interface level**, where the underlying data and AI algorithms, as also considered for AI more broadly, need to be examined. A key tenet of their account is that to develop a deep understanding of the ethical quality of AIED systems requires one to examine AIED through multiple lenses at those two levels. It is only by adopting such multiple lenses that a balanced understanding of the ethics of AIED approaches and systems can be achieved. We now briefly review the key factors pertaining to the two levels identified by Holstein and Doroudi, and through illustrative examples from AIED we elaborate on how this overarching framing of AIED's ethics aligns with the broader Ethical AI considerations reviewed in the earlier sections. As well as allowing to situate the ethics of AIED within the broader ethical AI debates, such an alignment is needed to identify specific areas where AIED can help advance those debates.

## 4.1 Equity of access to advanced technologies in AIED

The *digital divide*, between those who have access to suitable technologies and those who do not, has been the subject of research in education and of ongoing discussions within policy and media since the first wave of the ICT in Education in the early 2000s. The digital divide has been formally defined as "the gap between individuals, households, businesses and geographic areas at different socioeconomic levels with regard both to their opportunities to access ICT and to their use of the Internet for a wide variety of activities" (OECD, 2001, p. 4). In other words, certain groups do not have suitable access to ICT – a pre-requisite for being able to use AI applications. This is a disparity which holds implications for people's quality of life and the opportunities that are available to them. In this context, cautious observers of the recent drive to introduce and promote AI for Education have reasoned that in the best interests of students and teachers, before determining whether AI is of benefit to schools, it is necessary first to obtain rigorous evidence of both qualitative and quantitative dimensions of AIED (Facer and Selwyn, 2021; Holmes, Balik & Fadel, 2019; Nemorin, 2021). Central to these discussions is the question of how AI might exacerbate the digital divide and social inequities, rather than close the gaps, with some suggesting that AI in education may potentially result in amplified advantages for the individuals and organisations that have the ability to capitalise on it, while it disadvantages those who do not possess the requisite skills to use AI effectively (Carter, liu & Cantrell, 2020). The digital divide is typically determined by socio-economic factors, deriving from a combination of representational and allocative biases. In the context of education, where technology is increasingly incorporated as an essential tool for learning, and with education being a compulsory component of people's lives from early years to adulthood, the harms that might result from such biases being reinforced through AIED are likely to be profound and long-term.

The challenges related to equity of access to education are not limited solely to socio-economic factors or even to specific countries, or world regions. This was put in sharp focus during the COVID-19 pandemic, when in many contexts online learning became the norm and where many students were periodically excluded from education because they lacked adequate equipment or access to the Internet at home, even in technologically enabled countries. Lack of up-to-date or sufficient quantity of equipment is also a known issue in many schools in many countries (Baker, 2019). This leads to some schools being disadvantaged not only in terms of whether their pupils can access diverse technologies, but also whether their student populations are represented in research that powers 'computerised' or AI interventions' designs.

The unLOCKE project (Gauthier, Porayska-Pomsta & Mareschal, 2021; Gauthier et al., 2022; Wilkinson *et al.*, 2019) offers a recent and quite common example. This project aimed to ascertain the efficacy of a computerised neuroscience intervention to support primary school children learning of how to inhibit impulsive incorrect responses to counterintuitive maths and science problems. Substantial intervention and deployment challenges were experienced by the researchers owing to a large proportion of schools, mainly in the rural areas, not having adequate equipment or reliable Internet access. This determined how the intervention was delivered and limited what kind of data could be collected, which prevented the researchers gaining valuable access to the diversity of children's experiences and interaction patterns necessary to inform adaptive components of the system. While justified, this raises questions regarding representational quality of the data collected and illustrates how representation bias often arises and persists in AIED research. Other studies related to the impact of children's access to technology suggests that children with desktop or laptop computers access are more likely to use technology for learning (52%) than those who have mobile only access (35% of children) (Holstein & Doroudi, 2022; Rideout & Katz, 2016). Furthermore, stakeholders' (teachers, parents, school administrators) familiarity with and access to diverse ICT also impacts on whether there is the skill, the daring, and the appetite for innovative technology-mediated pedagogies.

### 4.2 Equity of Accessibility of AIED systems

*Access* to pre-requisite equipment is not the only source of ethical concerns in this context. Holstein and Doroudi highlight an important point related to linguistic and cultural *accessibility* of technology. In their examples of the under-explored barriers to equity of educational outcomes, they highlight the dominance of the mainstream language communication models and cultural references in the content (usually American or British English) of many AIED systems. Such models and references, when used with learners from non-dominant backgrounds, non-native speakers, or vernacular language speakers (e.g., African American English Vernacular: AAEV) likely hinder the accessibility of such technologies and reduce any resulting benefits. Indeed, Finklestein (2013) demonstrated that students using AAEV as their main language are likely to show greater scientific reasoning when interacting with technology that communicates with them in AAEV than in mainstream American English, suggesting that an investment in environments that are culturally and linguistically aligned with the learners may be essential to removing certain barriers to learning and academic achievement.

A broader point that relates to the culturally and linguistically inflexible designs of AIED is how learners' familiarity with the content presented affects their learning (e.g., maths problems situated in contexts familiar to students), the way in which such content might be delivered (e.g., through linguistically heavy word problems in maths vs. abstract equations), and specific learning methods (e.g., individual vs. collaborative learning methods). Culture-comparative neuroscience in education research (e.g., Ngan Ng & Rao, 2010; Tang & Liu, 2009) suggests that how we learn is not just related superficially to our specific cultural and linguistic traditions, but that it is also reflected in how we engage with learning content at a deep neuro-cognitive level. For example, fMRI studies provide emerging evidence of the differences between Western (*sic* American) and Eastern (*sic* Chinese) students' mathematical problem-solving. These differences are often conditioned by culturally determined ways of teaching mathematics in the USA vs. China and by the influence of language (in particular, number words in Chinese vs. English) on student number conceptualisation. In turn, how students conceptualise numbers has a knock-on effect on the efficacy of their mathematical problem solving. With this emerging knowledge and culture comparative evidence in mind, it is important to interrogate the questions about what exactly AIED systems ought to adapt, to whom they should adapt, and how they might create environments aligned with students' cultural and linguistic backgrounds.

Related to questions about how AIED adaptive support might contribute to more culturally inclusive education are questions about AIED inclusive practices in the context of physically disabled and neuro-diverse learners (or more generally – *differently abled* learners). There is a notable paucity of effort within AIED in this context, with a vast majority of AIED systems being geared towards mainstream education. This not only constitutes a pronounced representational bias in the field, but it also occludes important opportunities for innovation. For example, looking through the perspective of human-computer interaction design, Treviranus (2022) observes that research focused on the 'norm' rarely can be transferred to those outside the so-called 'norm', whereas going in the opposite direction is relatively straightforward. The emphasis on the so-called norm, in turn, proliferates exclusive practices within education and representational harms of recognition, under-representation and ex-nomination (see earlier sections). The few examples of AIED systems that focus on neurodiversity (e.g., Rashedi et al., 2020; Porayska-Pomsta et al., 2018; Benton et al., 2021) suggest that AIED's relative lack of emphasis on non-mainstream education is a missed opportunity not only with respect to enhancing AIED's inclusive practices, but also for leveraging the affordances of digital technologies and insights from special needs education to enhance the adaptive capabilities of AIED systems. This is particularly important for the deployment of AIED systems in a way that is relevant to users in specific contexts, and to making a real and positive difference for a diversity of learners (Porayska-Pomsta & Rajendran, 2019). For example, based on the ECHOES project (Porayska-Pomsta et al., 2018), which focused on developing AIED support for autistic children's social communication skills acquisition via an adaptive virtual peer, the SHARE-IT project (Porayska-Pomsta et al., 2013) investigated how the AIED system's interaction and pedagogy may be enhanced by allowing parents and teachers to modify user and communication models of the ECHOES AIED system. The findings suggest an acute need for technologies that are not pedagogically prescriptive, which give educational practitioners the flexibility to change diverse parameters such as timings and types of feedback given by the system, and which allow for *shareability* of the models between stakeholders to enable adaptation that is bespoke to individual learners.

The idea of modifiable models, as discussed above, is aligned with the ethos of editable and negotiable open learner models (OLMs; e.g., Bull and Kay, 2016), with the difference being that the users doing the editing are educators and caregivers. The insights here, which may be easily missed in mainstream contexts of education, are two-fold. The first insight relates to the role that AIED systems can play in fostering the development of shared support mechanisms for individual learners that rely on in-depth negotiation between different environments inhabited by the learners, e.g., between home and school. Through negotiation, facilitated by the specialised AIED approaches such as OLMs, educators and caregivers are better enabled to understand learners' needs across diverse contexts and to deliver wholistic support. In extreme cases this can lead to a complete overhaul of the approaches adopted with the individual learners, since the behaviours manifested at home may be sometimes or completely absent in classroom environments (e.g., the willingness of a child to communicate; Bernardini, Porayska-Pomsta, Smith & Avramides, 2012). Thus, understanding learners in diverse contexts is often necessary for creating entirely new support environments (physical and pedagogical) in which they can build on their strengths rather than be identified by their deficits (Guldberg, Keay-Bright, Parsons & Porayska-Pomsta, 2017). The second related insight spotlights the need for flexible designs and open models that can be modified by education experts and caretakers *on-demand* and according to the changing circumstances and needs of the students. Such changing needs and circumstances may be inaccessible to even the most sophisticated AIED systems. Naturally, creating such flexible designs does present, substantial technical challenges, but it also raises a possibility of creating AIED systems that are more attuned with their users, that inspire and aid improvements in the wider system of education, and that more readily reflect the field's aspirations to improve educational and life outcomes for all. Indeed, such flexible designs also align with an emergent vision of responsible AI more broadly – a vision that challenges the ethical and societal value of the standard model on which AI systems are presently based (Russell, 2019; Porayska-Pomsta and Holmes, 2022).

## 4.3 Equity of AIED's pedagogies

Linked to the learning support innovations that may be missed because of representational biases in AIED's research is the issue of the dominant pedagogies that are promoted through AIED's systems. Driven by the ambition to achieve the 2-sigma effect identified by Bloom and his colleagues in the context of one-to-one mastery learning, AIED as a field has disproportionately concentrated its efforts on drill and practice type of pedagogies, and exam-type assessments. This, combined with the relatively easier to formally model and assess well-defined science subjects, gives rise to forms of bias that are very specific to the context of education. We propose to call these biases the *domain-value bias* and *learning-culture bias,* respectively. The **domain-value bias** refers to the society valuing some subject domains more than others. Arguably mathematical and biological sciences are presently considered more valuable than arts subjects. This is acutely visible in the national curricula across much of the world, where subjects such as history or arts are no longer obligatory, while other subjects such as philosophy, are poorly subscribed to at universities.

We use the term **learning-culture bias** to refer to some modes of learning tending to be more dominant and more *bankable* in the mainstream educational contexts than others. For example, drill-and-practice, exam-oriented type of learning tends to dominate in many mainstream educational contexts across the world as it forms part of a wider educational machinery that requires educational institutions' administrators and politicians to quantify (usually in monetary terms) people's intellectual and societal worth. However, while in some circumstances the mastery-learning and exam-based pedagogies are of value to the development of people's skills and erudition, they often leave little room for curiosity-driven enquiry, exploration, discovery, collaboration, or for productive failure. In this, they contribute to the proliferation and entrenchment of an oppressive educational system, where increasingly, one can only be either an A* genius or an academic failure (e.g., see typical university entry requirements in the UK). While systems that employ OLMs or those that promote self-regulation and metacognitive skills in their pedagogies (e.g., Bull and Kay, 2016; Blair, Schwartz, Biswas, Leelawong, 2007; Aleven et al., *in this volume*) implicitly mitigate some of this bias, it is important to interrogate whether

AIED's ethical value might be enhanced by the field's greater investment in going against the established grain within a broader educational system to offer more diverse and daring forms of learning supports.

Both the domain-value and learning-culture biases within the broader education systems are exploited and reinforced by the EdTech industry, much of which has seized on the current AI hype. While the extent to which many of the EdTech companies actually offer AIED-driven systems is debatable, it is hard to overlook the EdTech industry's efforts, encouraged by policy makers, to bank on the exam-driven, hard sciences favouring education system. In this context, Blikstein (2018) spotlighted a number of disturbing EdTech trends of pertinence to the ethics of AIED, which proliferate and reinforce questionable pedagogical practices. Specifically, he highlighted the persistent rhetoric surrounding EdTech and the policy's push towards embedding technologies in classrooms and the related funding in and out of academia (see also, Madaio, 2022; Ritter & Koedinger, *in this volume*). This rhetoric focuses on persistent, but broadly unevidenced claims of technologies' ability to rescue teachers from boring and repetitive tasks by making learning management information systems more efficient in order to allow teachers and administrators to allocate their time more efficiently to less mundane tasks (Miao & Holmes, 2021; Watters, 2021), and on technologies' ability to offer endless explanations to students, along with opportunities for repetition, drill-and-practice and breaking content into tiny pieces of information. He spotlighted the ethical risks associated with adopting this rhetoric, which stands in stark contrast with evidence-based educational practices that have debunked the transmitter model of education as both inefficient and oppressive (Blikstein, 2018). Ultimately, not only the rhetoric behind the educational technologies' industry fails to align with the best educational practices, but the industry's practices are also notorious for lack of any auditing procedures (du Boulay, 2022) that would ensure the safety of the systems for the learners, some as young as toddlers. Here, a key question for AIED as a field, relates to its potential role in guiding the best EdTech practices and highlighting both the opportunities for educational innovation, and in identifying and addressing related ethical challenges.

## 4.4 Interaction between human and machine autonomy

Important ethical considerations also arise because of context-specific interactions between humans and the AI artefacts. Holstein and Doroudi (2022) highlight the importance of appreciating the complexity of the socio-technological ecosystem, including the AI systems (the artefacts), the designers of the technology, and the users (decision makers, teachers, school, administrators, and learners). They emphasise the need to design carefully for the interplay between the different elements within this ecosystem to ensure more equitable futures for AI in Education. They observe that even if AIED or Learning Analytics do not contain harmful biases in their underlying algorithms or data, the way in which they are mediated at the user interface level, may still lead to potentially undesirable or harmful effects, through the systems' interaction with the users' pre-existing beliefs and habits. They suggest that it is critical to consider the intent with which a tool is built, and the human-AI interaction that is facilitated. They focus specifically on the example of teachers as users and propose that if a tool has the capacity to challenge users' *a priori* beliefs and ways of thinking, then it may help them develop and/or adopt more equitable practices. If a tool does not possess such capabilities, then it may only serve to reinforce pre-existing beliefs and associated biases. Ultimately, while humans shape the technology to their needs and aspirations, the technology inevitably also shapes the humans. A key conclusion delivered by Holstein and Doroudi is that to address any ethical challenges in AIED, the field ought to be centrally concerned with understanding and designing for such human-AI feedback loops.

The idea that we shape our technologies and are shaped by them in return is well rehearsed within the philosophy of technology and AI: from Kelly's concept of *technium* – a self-reinforcing system of technological inventions and reinventions (Kelly, 2010), through Kurzweil's ideas around technological singularity arising from ever-accelerating technological inventions (Kurzweil, 2014), to more recent discussions by Russell of the fundamental ways in which technology changes us (Russell, 2020), and Harrari's historically situated examination of the rise of biotech as means for human enhancement (Harrari, 2019). Although these perspectives may seem far away from the AIED work and focus, we argue that they in fact raise fundamental questions for the field and demand new lines of investigation that pertain to the impact of the AIED technologies on users' cognition, perception of their environment and decision-making capacities that go beyond subject-based assessment of knowledge. They put in sharp relief the possibility of trade-offs related to technological enhancements, e.g., scheduling systems that allow us to be more organised at the cost of lost opportunities for exercising working memory (The Royal Society, 2019); or use of learning analytics to aid assessment of students in quantifiable curricula, potentially at the cost of lost opportunities for teachers to rehearse and critically appraise their students' journeys as developing humans who inhabit complex socio-emotional environments (see also Arroyo, Muldner & Porayska-Pomsta, *in this volume*). Here the questions for the field go beyond the specific AIED systems that we build. They extend to the need to interrogate what role AIED as a field that is concerned with the interaction between AI and human cognition wants to and can play in influencing and shaping education of the future. These questions spotlight that the ethics of AIED are not only about addressing biases and harms within the systems we build, but they are also fundamentally about how the field contributes (actively or through complacency) to entrenching education as an exam-passing machine, and also potentially, to de-skilling and de-professionalisation of teachers by virtue of automating some of the most important of their tasks such as assessment of students – a task, demanding and time consuming as it is, that allows teachers to connect with their students as living, breathing, developing humans.

## 4.5 The ethics of datasets, algorithms, and methods in AIED

Many of the ethical considerations related to data, algorithmic bias and methods used in AI more generally apply in the context of AIED. Specifically, many of the same ethical challenges and questions that arise during the machine learning pipeline discussed in earlier sections, are of relevance to educational contexts. Issues of allocative and representational bias can be observed across AIED work with substantial gaps related to the diversity of demographic categories also being apparent. Earlier in this chapter we offered some examples of how representational bias may arise, e.g., due to limited access to schools with adequate equipment. Baker and Hawn (2021) provide a detailed inventory of groups of learners (categorised by demographics and protected characteristics) for whom there is little data in AIED research. Specifically, AIED datasets lack in diversity with respect to students' different ethnicities, nationalities, different-*ableness* (physical and neuro-cognitive), urbanicity, parental education, socio-economic status, international students, and military connected status. Inclusion or exclusion of any of these categories in the AIED designs may lead to disparities in the fairness of treatment by and outcomes from using an AIED system. In previous subsections, we discussed some of the representational biases in and out of the context of education. For a detailed inventory and analysis of the representational gaps in AIED research we refer the interested reader to Baker and Hawn (2021). In this section we instead focus briefly on the challenges related to algorithmic fairness in AIED and the methods that are available, or that we should be developing to mitigate bias in AIED systems. Understanding these challenges is important for formulating ethics-related questions that AIED researchers need to consider before, during, and after the design and deployment of their systems.

### *4.5.1 Algorithmic fairness in AIED*

Kizilcec and Lee (2022) and Baker and Hawn (2021), offer a couple of most recent overviews of issues pertaining to algorithmic fairness and related data collection methods in the context of AIED, drawing attention to the types of assumptions and methodologies that impact on the quality of data and algorithms in AI applications for education. Kizilcec and Lee elaborate on the challenges identified in the broader AI context with respect to defining fairness and building fairness into AIED algorithms. They highlight equity and equality as two somewhat contradictory central notions related to fairness in education, and they link those notions to questions about disparities between how AIED's diverse users are treated vs. how they are impacted by an algorithmic intervention. Specifically, while equality may be achieved through innovation if all individuals benefit the same amount regardless of their pre-existing capabilities, to achieve equity (i.e., closing the achievement gaps) the impact of innovation must be positively greater for those who start from behind. This positioning presents a set of questions, likely some dilemmas, and the obligation of transparency for AIED designers with respect to both what form of algorithmic fairness they choose to furnish their systems with and what claims they can make about the generalisability of their applications to diverse users and contexts.

Three representative notions of fairness are discussed by Kizilcec and Lee, each highlighting different and not necessarily compatible ways in which fairness may be considered. The first notion of fairness is **statistical fairness**, which relies on three fairness criteria: (a) *independence*, which is satisfied if an algorithm's decisions are independent of group membership; (b) *separation*, which is satisfied if the algorithm makes correct or incorrect predictions independently of a group membership; and (c) *sufficiency*, which is satisfied if an algorithm's decisions are equally significant for all groups. All three criteria can be useful in enhancing algorithmic fairness, but all three can also lead to unpredictable outcomes, depending on the context and purpose of the algorithm's application (see also Baker and Hawn (2021) for a discussion of the trade-offs between these different fairness metrics and fairness of outcomes).

Approaches to the second notion of fairness, **similarity-based**, are known as group measures of fairness, because they ignore individual features of and differences between cases. Typically, fairness criteria for similarity-based approaches include *fairness through unawareness*, i.e., achieved by ignoring during model training any protected attributes in the data; and *individual fairness*, which involves constructing similarity metrics between individuals for specific prediction tasks. While evaluations of the fairness through unawareness approach suggest improvements in accuracy in some decision-making contexts such as algorithmic admissions systems (Kleinberg, Ludwig, Mullainathan & Sunstein, 2018; Yu, Li & Wu, 2020), they also carry a danger of a model inadvertently reconstructing the protected attributes from features which may seem unrelated. The individual fairness metric addresses this issue, but its specific weakness relates to its reliance on distance metrics (which may also contain fairness imbalances) and on the assumption that treating similar individuals similarly will lead to the same outcomes.

The final notion of fairness discussed by Kizilcec and Lee is the **causal notion of fairness**, whereby an algorithm can be considered fair if it produces the same predictions under different counterfactual scenarios, e.g., predicting the same outcome for an individual by varying a specific feature, e.g., gender, while keeping all other known features constant. The idea behind this approach derives from the observation that understanding how different predictions might change for different group memberships relies on causal inference. While offering a way to evaluate equality of predictions, the causal notion of fairness relies on the validity of a causal model, which needs to make predictions based largely on observational data. Therefore, this approach is itself open to incorporating diverse biases.

While the above three notions of fairness and the corresponding approaches represent only a subset of existing conceptualisations of algorithmic fairness, they demonstrate the difficulty of finding a principled way of selecting metrics of fairness and indeed, of measuring fairness in a way that satisfies all possible scenarios of AIED applications. In this context, Kizilcec and Lee suggest that more than one approach may be needed to evaluate fairness for each AIED scenario. Such an approach would promote concrete discussions about the ethical value of any specific AIED system. However, to facilitate such discussions demands from AIED designers' clarity and transparency of goals related to equity and equality that their specific systems aim to deliver.

### 4.5.2 Methodological considerations for Ethical AIED

Fundamental to the choices of fairness metrics are also questions about data collection methods used in AIED. Baker and Hawn (2021) draw attention to the critical role of data collection choices in mitigating diverse forms of algorithmic bias downstream in system development and deployment. Specifically, they focus on two key forms of data bias in AIED, namely representational bias and measurement bias, and on the corresponding methods that may lead to or mitigate against those biases. Addressing representational bias in data is fundamental to developing AIED systems that cater for diverse learners and contexts of use. Achieving representationally balanced data is not trivial and cannot be guaranteed by simply adopting a proportional sampling method. Although, Baker and Hawn recommend that over-sampling is better than under-sampling, they also highlight real world obstacles in this context, such as the tendency of only certain groups of participants (e.g., learners in privileged socio-economic urban areas) being available to take part in data collection efforts, lack of appropriate infrastructure in schools or access to technology at home, as well as research-readiness of institutions and individual decision makers responsible for promoting or facilitating links between institutions and academic research.

Following Suresh and Guttag (2020), earlier in this chapter, we defined measurement bias as potentially arising in training labels and predictor variables. Baker and Hawn contextualise this for some of the most common methods employed in AIED research. They are mainly concerned with the measurement bias that arises in training labels and they provide a number of indicative examples that are pertinent to AIED. They include in their list labelling bias that arises from (a) historical prejudices such as racism which might discriminate against specific populations (e.g. students of colour); (b) mismatch between the culture of the coders and the culture of the learners whose data is being encoded (e.g., when coding emotions based on facial expressions), and (c) self-reporting (in which learners may carry a bias towards self, due to diverse possible factors such as lack of confidence, cultural factors, or fear of being stereotyped or judged).

While Baker and Hawn are less clear on the subject of predictor bias, it is also important to consider how such bias may influence the ethical value of an AIED system, in particular whether it offers a reasonable proxy for the kind of thing that the system is trying to predict. This is particularly important in the context of student modelling where interaction data with the system is often used to predict things such as student understanding of a topic, based on their performance on specific problems, or level of interaction with the system. In this context, the appropriateness and validity of the predictors chosen have to be examined in the context of the pedagogies that an AIED aims to encapsulate. For example, if an AIED system's purpose is to train students to perform well on particular types of problems, then while the students may do well on such problems, they may still not have a good knowledge of the other related elements of the domain within which these problems exist.

It is important to acknowledge that, while many ethics-related questions need to be considered at every step in an AIED system development, in many ways, as a design science, AIED has been ahead of other subfields of AI in the adoption of methods that connect AIED systems and algorithms with humans who use them. There are many exemplars of user-centred and participatory design approaches across different AIED projects (e.g., Grawemayer et al., 2017; Porayska-Pomsta et al., 2018), with knowledge elicitation methodologies and contextualised interaction data forming the backbone of many AIED models and decision-making processes. In this, AIED has always aimed to be not only relevant to the users, but also educationally efficacious, and AIED researchers have long understood the importance of the contextual and representational validity of data, and of the methods used to generate such data (Porayska-Pomsta et al., 2013). Open Learner Models (OLMs) are in many ways unique to AIED insofar as they offer the users a degree of ownership over their data, explicitly acknowledging that AI models are not accurate and that AI *transparency* and *explainability* plays a pivotal role in supporting human learning and criticality (Bull and Kay, 2016; Kay et al., *in this volume*; Conati et al., 2019). In a number of ways OLMs may represent the future for AIED – for increasing the fields ethical value in its practices and solutions – a future which also aligns with Holstein et al.'s (2019) recommendations for how to address the challenges of bias and how to improve on fairness of AIED by allowing the educational practitioners to collect data themselves.

## 5. Towards a framework of Ethics of AIED

In this chapter, we have emphasised the need for any effective ethics of AIED to be robustly grounded in broader ethical debates – specifically, the ethics of AI in general, in moral philosophy, and in the ethics of educational practice. Yet, as we have noted, while a large number of ethical principles are frequently invoked for AI in general, it is not always clear how those principles are best enacted in the particular domain in which the AI is being applied. While issues such as the ethics of data use (e.g., consent and data privacy) and the ethics of models and algorithmic computations (e.g., transparency and fairness) are transversal – they have relevance for all domains in which AI is applied – the AI domains of healthcare, transport, ecology, education, and others self-evidently all have domain-specific ethical issues that also need to be properly considered and addressed (Holmes et al., 2021). Ethical challenges in education that need to be thoroughly considered include: the accuracy and validity of assessments, what constitutes useful knowledge, educators' roles, and agency in selecting pedagogies that suit their learners' needs best, and particular approaches to pedagogy.

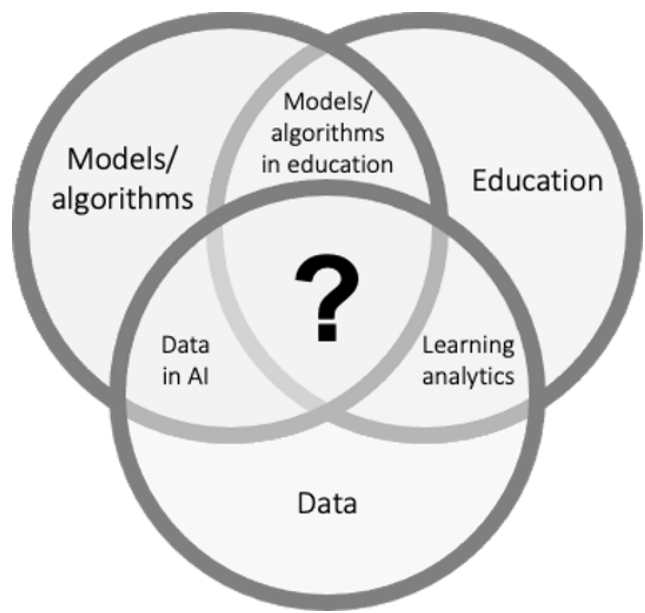

*Figure 1 'Strawman' framework for the ethics of AIED, based on Holmes et al (2021)*

To help AIED practitioners conceptualise and address the ethics of AIED in their work, elsewhere we hypothesised a 'strawman' draft framework (Holmes et al., 2021) (Fig. 1). This framework, which begins with the three ethics foci identified (data, models, and education) was designed for the purpose of stimulating discussion. In this framework, data, models, and education constitute the foundational level; while at the overlaps is a second level: the ethics of data in AI (e.g., Floridi and Cowls, 2019), the ethics of data in education (e.g., Ferguson *et al.* 2016), and the ethics of models and algorithms applied in educational contexts; the last of which remains the least developed area of research. There is also the central intersection: the specific interaction between AI systems and human cognition at the individual level (indicated by the question mark in Fig. 1).

This framework was proposed as a first step in the necessary conversation and as such it is open to refinements by insights from all the emergent work in this area. Here, we propose a second step, by recognising and disambiguating two fundamental dimensions of the broader concept of education: (1) pedagogy, or educational practices, and (2) the socio-technical context within which education happens. In Table 2, we explore this amended framework (comprising four foci: 'Data', 'Models/Algorithms', 'Pedagogy', and the 'Socio-technical context') in terms of the different biases identified throughout the chapter. For each, we offer an example problem and necessary (but by no means exhaustive) set of questions that AIED practitioners may need to consider and address. How we as a community formulate and address such questions may have profound existential implications for how our societies and our cultures progress and develop, and our individual social, cognitive and physical functioning changes. Accordingly, for each set of problems and questions, we also offer an indicative set of consequences or risks associated with the diverse biases. This approach aims to avoid the framework becoming or being seen as only generating a list of prohibitions to stop 'unethical' AI in education activities. Instead, the aim is to provide a foundation to facilitate forward thinking and *ethics-conscious AIED research and development by design*.

*Table 2 Key Considerations & Questions for AIED designers and users*

| | **DIMENSIONS OF AIED DESIGN AND DEPLOYMENT** | | | |
|---|---|---|---|---|
| | **AIED Data** | **AIED Models/Algorithms** | **AIED Pedagogy** | **Socio-technical context** |
| **Historic** | Problem: | | | |
| | • Historic datasets harbouring prejudices or stereotypes (e.g. boys are good at sciences, girls are good at arts) | • Pre-existent prejudices or inflexible pedagogies encoded in the choice of an objective function or heuristics (e.g. language models based on mainstream language forms such as mainstream American or British English for learners who speak vernacular English) | • Inflexible or limited or outdated pedagogies such as knowledge transmission, drill and practice approaches | • Students from lower socio-economic backgrounds or minorities having lower access to resources<br>• Education as a mechanism for transmitting information, and training students to pass exams<br>• Devaluation of soft skills and subject domains |
| | Questions: | | | |
| | • How old is your data?<br>• Where does it come from?<br>• Have you checked for prejudices and stereotypes therein? | • Who are your target learners and contexts of use?<br>• Have you checked that your assumptions are adequate for those learners and contexts? | • Is the choice of pedagogy adequate for the educational goals you aim to achieve?<br>• How flexible is your system in allowing users to modify the pedagogical model? | • Is your system affordable and usable across contexts?<br>• Have you accounted for the infrastructure available in the target context?<br>• Is your system reinforcing or innovating existing educational practices? |
| | Consequences/Risks: | | | |
| | • Harm of representation<br>• Outcome fairness<br>• Equity of access | • Harm of representation<br>• Outcome fairness<br>• Equity of accessibility | • Harm of representation<br>• Equity of accessibility<br>• Outcome fairness<br>• Process Fairness<br>• Human autonomy | • Harm of allocation<br>• Harm of representation<br>• Outcome fairness<br>• Equity of Access<br>• Equity of Accessibility |
| **Representational** | Problem: | | | |
| | • Sampling is not representative of student populations (e.g. data contains gender, race, urbanicity bias etc.). | • Models that generalise assumptions (objective function, or inference) reflecting limited population subsets or designers/software developers' pre-conceptions. | • Prescriptive pedagogies which accommodate and enforce limited learning cultures (e.g. exam-oriented drill and practice). | • AIED aims not being representative of educators' aims or diverse learners' needs. |
| | Questions: | | | |

| | DIMENSIONS OF AIED DESIGN AND DEPLOYMENT | | | |
|---|---|---|---|---|
| | **AIED Data** | **AIED Models/Algorithms** | **AIED Pedagogy** | **Socio-technical context** |
| | • Who are the target populations for your system?<br>• Is your data representative of all students and contexts in which you envisage your AIED, some, or only one specific context? | • How are your models' assumptions validated?<br>• Do your choices of objective function/inference align with diverse learners' behaviours, cultures, and forms of communication? | • Is your AIED able to accommodate different learning cultures and needs of the users?<br>• How flexible are your pedagogical and communication models? | • Does your system respond to a diversity of needs of educators and/or learners?<br>• How did you establish those needs (e.g. through literature, by working with diverse stakeholders, etc.)? |
| | Consequences/Risks: | | | |
| | • Harm of representation<br>• Equity of accessibility<br>• Outcome fairness | • Harm of representation<br>• Equity of accessibility<br>• Outcome fairness<br>• Process fairness | • Harm of representation<br>• Equity of accessibility<br>• Outcome fairness<br>• Process fairness<br>• Human autonomy | • Harm of representation<br>• Equity of accessibility |
| **Measurement** | Problem: | | | |
| | • Sampling is not representative of student populations (e.g. data contains gender, race, urbanicity bias etc.).<br>• Sampling is inconsistent between contexts<br>• Mismatch between data coders' and data donors' cultures (e.g. Americans coding emotional states of Turkish students). | • Inappropriate choice of a predictor variable (e.g. conflating login-logoff duration with engagement or clicking a document download link with reading the document).<br>• Objective function or inference mechanism based on. | • Mismatch between learners needs and the support they receive (e.g. neuro-divergent learners receiving support designed for neuro-typical learners including inappropriately timed or volume of feedback, over-reliance on reading instructions, etc.).<br>• Misinterpretation of learner behaviours, leading to inaccurate or unfair assessments. | • Reliance on data from dominant groups to devise pedagogies and school/district/national educational policies.<br>• Entrenchment of limited forms of assessment.<br>• De-skilling educators by introducing process and outcome measures (e.g. through LMSs, which may be inaccurate, inadequate, or meaningless to educators (click data as a measure of engagement, response on specific test problems as a measure of student knowledge). |
| | Questions: | | | |
| | • Have you checked for proportionality in your sample vis a vis the target population?<br>• Have you controlled for the consistency of application of methods and procedures across contexts (e.g. by training and | • Did you verify the pedagogical and cultural validity of your predictors (e.g. using student click data to predict engagement, or norms of one culture to predict student emotions of another culture)? | • Is your system's pedagogical approach appropriate for the target group?<br>• Does/should your system allow for any adjustments to be made (either a priori or in-real time) by human practitioners (e.g. to improve the idiosyncrasy in your | • Is your system reinforcing or challenging the dominant groups education culture?<br>• Is your system challenging, enhancing, or confirming educators' prior beliefs and assessments? |

| | DIMENSIONS OF AIED DESIGN AND DEPLOYMENT | | | |
|---|---|---|---|---|
| | **AIED Data** | **AIED Models/Algorithms** | **AIED Pedagogy** | **Socio-technical context** |
| | auditing researchers collecting data in different schools)?<br>• Is there cultural alignment between data coders and data donors? | • How did you verify the validity of your predictors (e.g. by following other researchers, by exploring with and eliciting from experts)?<br>• Can users negotiate with/ change your models (e.g. through OLM; adjusting/ choosing communication model, etc.)? | system's diagnoses) of diverse learners? | • Does your system enhance or undermines human educators' skills and practices by offering insights that meaningful to them (e.g. by reference to process and outcome measures that are familiar to teachers /grounded in best educational practices) |
| | Consequences/Risks: | | | |
| | • Harm of representation<br>• Outcome fairness<br>• Process fairness | • Outcome fairness<br>• Process fairness<br>• Human autonomy | • Harm of representation<br>• Equity of accessibility<br>• Outcome fairness<br>• Individual fairness<br>• Human autonomy | • Harm of representation<br>• Equity of accessibility<br>• Outcome fairness<br>• Process fairness<br>• Human autonomy |
| **Aggregation** | Problem: | | | |
| | • Erasure of some groups or individuals from data (e.g. because of insufficient numbers).<br>• Demographic and protected categories blindness. | • Reduced adaptivity and relevance of systems to 'outlier' students due to models operating on dominant behavioural patterns. | • Inappropriate, discriminatory, or ineffective support for the target group (e.g. pedagogies for non-dyslexic used for dyslexic students). | • Reliance on aggregated/ historic data to derive assessments for individuals (e.g. 2020 A-level fiasco in the UK).<br>• Lack of educationally inclusive practices; practices that are blind to non-typical, non-dominant groups (e.g. neurodivergent, low SES, or protected categories students). |
| | Questions: | | | |
| | • Have you assessed the risks of aggregating data to your systems' ability to adapt to your target population?<br>• Have you assessed pros and cons of data aggregation vis a vis the goals of your system (e.g. university admission vs adaptive support). | • What is the necessary and sufficient level of the representational granularity of data for your system to model and adapt effectively to individual or groups of learners? | • Are the pedagogies encoded in your system based on best practices available for the target population?<br>• Can your system's pedagogy be adapted by human educators or through negotiation with learners? | • Does/can your system actively contribute to increasing individual fairness?<br>• Does your system encode inclusive educational practices and if so, how? |
| | Consequences/Risks: | | | |

|  | DIMENSIONS OF AIED DESIGN AND DEPLOYMENT | | | |
|---|---|---|---|---|
|  | **AIED Data** | **AIED Models/Algorithms** | **AIED Pedagogy** | **Socio-technical context** |
| **Model Learning** | • Harm of representation<br>• Harm of allocation<br>• Outcome fairness<br>• Process fairness<br>• Individual fairness<br>• Group fairness | • Harm of representation<br>• Harm of allocation<br>• Outcome fairness<br>• Process fairness<br>• Individual fairness<br>• Group fairness | • Harm of representation<br>• Harm of allocation<br>• Outcome fairness<br>• Process fairness<br>• Individual fairness<br>• Group fairness<br>• Human autonomy | • Harm of representation<br>• Outcome fairness<br>• Individual fairness<br>• Group fairness |
|  | Problem: | | | |
|  | • Data which is too small and/or not representative enough of the target population or problem to facilitate model learning or learning of model that generalises across contexts.<br>• Relying on inadequate predictor variable (e.g. click data as predictor of underlying intentions behind a particular search term). | • Choice of an objective function, heuristics or inference mechanism that leads to performance differences across different data.<br>• Prioritising one objective over another may limit the applicability of the model to different learner populations and contexts of use. | • Pedagogical choices based on the model may not transfer across different contexts (e.g. the same help-seeking behaviours may be explained differently depending on the target population, their learning culture, neuro-divergent status, etc.)<br>• Users' decisions and actions may be adversely affected by predictions/inferences or recommendations that are incorrect, irrelevant, or based on incomplete information. | • A model that does not generalise across contexts may reinforce/introduce exclusive or harmful practices in education.<br>• A model that makes incorrect, inaccurate, or irrelevant predictions or recommendations may influence users behaviours in undesirable ways (e.g. it may compromise the quality and fairness of educators' pedagogical decisions). |
|  | Questions: | | | |
|  | • Does the volume and representational quality of your data sufficient to cater for diversity of contexts?<br>• What evidence have you chosen your predictor variables based? | • To what extent does your chosen objective function/heuristics/inference mechanism allow your model to make meaningful and pedagogically appropriate predictions/recommendations/action? | • To what extent does your objective function/ heuristics/ inference allow your model to account for users' idiosyncratic behaviours and needs?<br>• Does your model promote users' critical thinking (e.g. by inviting them to challenge the models' predictions/ recommendations), or does it dictate the 'answers'?<br>• Is your model explainable to all/some the users? | • What contexts, what users and what educational tasks does your model support?<br>• How well does it generalise to new contexts, users, and tasks?<br>• Does your model dictate a particular way of analysing the users or supports a specific mode of learning and teaching, or is it open to being critiqued and modified? |
|  | Consequences/Risks: | | | |

|  | DIMENSIONS OF AIED DESIGN AND DEPLOYMENT ||||
|  | AIED Data | AIED Models/Algorithms | AIED Pedagogy | Socio-technical context |
|---|---|---|---|---|
| **Evaluation** | • Harm of representation<br>• Harm of allocation<br>• Outcome fairness<br>• Process fairness | • Harm of representation<br>• Harm of allocation<br>• Outcome fairness<br>• Process fairness<br>• Individual fairness<br>• Group fairness | • Harm of representation<br>• Harm of allocation<br>• Outcome fairness<br>• Process fairness<br>• Individual fairness<br>• Human autonomy | • Harm of representation<br>• Harm of allocation<br>• Outcome fairness<br>• Process fairness<br>• Individual fairness<br>• Human autonomy |
|  | Problem: ||||
|  | • Benchmark data may fail to match the user population data. | • Model performance might be evaluated using different performance metrics, each metric presenting its own challenges (e.g. aggregate measures can hide subgroup underperformance).<br>• Overfitting of a model performance based on comparison with other models against the benchmark data. | • Pedagogical decisions/recommendations which are ineffective, irrelevant, or even harmful. | • Overgeneralised claims about the efficacy and relevance of a given technology.<br>• Choice of model performance metrics may cater for the needs and interests of some stakeholders (e.g. businesses, policy decision makers), but not others (e.g. students, front-line practitioners).<br>• Investment in pedagogically limited, prescriptive, or potentially harmful technologies. |
|  | Questions: ||||
|  | • How representative is your benchmark data of the user population? | • What performance metrics are needed to ensure educational efficacy, relevance, and safety of your models?<br>• What have you compared your model's performance to? | • Are your system's models representative of all target users in all target contexts?<br>• In what way exactly are your system's pedagogies efficacious?<br>• What are the pedagogical limitations of your system?<br>• How transparent are the limitations of your system to the users? | • Who is the evaluation for (e.g. engineers, educators, educational administrators, investors, policy decision-makers)?<br>• How transparent are the limitations of your system to different stakeholders? |
|  | Consequences/Risks: ||||
|  | • Harm of representation<br>• Harm of allocation<br>• Outcome fairness<br>• Process fairness | • Harm of representation<br>• Harm of allocation<br>• Outcome fairness<br>• Process fairness<br>• Individual fairness | • Harm of representation<br>• Harm of allocation<br>• Outcome fairness<br>• Process fairness<br>• Individual fairness | • Harm of representation<br>• Harm of allocation<br>• Outcome fairness<br>• Human autonomy |

|  | DIMENSIONS OF AIED DESIGN AND DEPLOYMENT | | | |
|---|---|---|---|---|
|  | **AIED Data** | **AIED Models/Algorithms** | **AIED Pedagogy** | **Socio-technical context** |
| | | • Group fairness | • Group fairness<br>• Human autonomy | |
| **Deployment** | colspan Problem: | | | |
| | • Data (learning, test, or benchmark) is not collected in deployment context(s) | • Decontextualised model designs and evaluations.<br>• Inflexible model designs that exclude the users from calibrating them in-situ. | • Learning designs for homogenous group and mainstream contexts do not transfer to non-typical groups (e.g. pedagogies used for neuro-divergent students usually transfer to neuro-typical students, but not the reverse). | • Socio-technical context ignored in the evaluation of your system leading to mismatch between the problem your system trying to solve and the way it is used. |
| | colspan Questions: | | | |
| | • Does the design context match the context of deployment?<br>• Is your data collected in representative contexts of use?<br>• Can the context in which the system is deployed change? | • Do your system's models encode the contextual information needed for their performance efficacy and relevance to those contexts? | • Can your system's pedagogy be changed, adjusted, or 'sabotaged' by human practitioners should they need to calibrate it to their specific needs (e.g. through adjusting when and how to deliver feedback to specific learners)? | • In what context is your technology deployed?<br>• Does your system design context match the context(s) it claims it will work in?<br>• Is your system transparent about its strengths and weaknesses?<br>• Can your system's use be modifying by front-line users? |
| | colspan Consequences/Risks: | | | |
| | • Harm of representation | • Harm of representation<br>• Outcome fairness<br>• Process fairness | • Harm of representation<br>• Harm of allocation<br>• Outcome fairness<br>• Process fairness<br>• Individual fairness<br>• Group fairness<br>• Human autonomy | • Harm of representation<br>• Harm of allocation<br>• Outcome fairness<br>• Process fairness<br>• Individual fairness<br>• Group fairness<br>• Human autonomy |
| **Domain-value** | colspan Problem: | | | |
| | • Datasets available in limited types of subject domains (e.g. sciences), with sparse data being available in other domains<br>• Subject domain may interact with performance of students belonging to different | • Models specialised for limited types of subject domains.<br>• Models for less favoured domains built on the assumptions (objective function etc.) derived from research in favoured domains (e.g. student | • Unequal access to AIED resources for educators in different domains (e.g. maths vs history)<br>• Danger of segregation of skills (e.g. problems solving on | • Investment in AIED supports for the more valued domains (e.g. maths and sciences)<br>• Generalisation of claims made for a limited set of AIED models and technologies to the whole of education. |

| | **DIMENSIONS OF AIED DESIGN AND DEPLOYMENT** | | | |
|---|---|---|---|---|
| | **AIED Data** | **AIED Models/Algorithms** | **AIED Pedagogy** | **Socio-technical context** |
| | demographic groups, proliferating historic and representational biases and harms. | emotion modelling in maths applied in history learning context). | specific tasks vs creative and critical thinking).<br>• Limited pedagogical innovation. | |
| | Questions: | | | |
| | • Are sufficient and representative datasets available in your target domain? | • Do your system's models differentiate between subject-domain specific and subject-independent assumptions that drive the selection/definition of its objective function/inference? | • Is your system designed for a currently favoured or less favoured domain?<br>• Does your system's pedagogy draw from other domains explicitly to promote and build on diversity of students' skills (e.g. problem solving, creative and critical thinking)? | • Is your system addressing a gap in AIED support provision or fitting into/exploiting a trend? |
| | Consequences/Risks: | | | |
| | • Harm of representation<br>• Harm of allocation<br>• Outcome fairness | • Harm of representation<br>• Outcome fairness<br>• Process fairness | • Harm of representation<br>• Harm of allocation<br>• Outcome fairness | • Harm of representation<br>• Harm of allocation<br>• Outcome fairness |
| **Learning-culture** | Problem: | | | |
| | • Datasets available in limited cultures of learning (e.g. drill-and-practice, collaborative, enquiry, or project driven).<br>• Learning culture may interact with performance/outcomes of student demographic, cultural and neuro-ability groups, proliferating historic and representational biases and harms. | • System models specialised to specific forms of learning may be misaligned with the strengths and moment to moment needs of the students (e.g. a student may underperform in drill-and-practice context, but thrive in an enquiry one). | • AIED's support of learners and teachers may be limited and/or limiting to them.<br>• AIED may contribute to entrenching modes of learning (e.g. exam-driven drill and practice) as dominant. | • Pedagogies that lead to examinable outcomes dominate the education system.<br>• AIED is expected to support the current policies and modes of learning dominant within the current education system. |
| | Questions: | | | |
| | • Are sufficient and representative datasets available in your system's target learning mode?<br>• Have you assessed for strengths of target learners in the target learning mode (e.g. some | • Do your system's models encode a single/specific mode of learning (e.g. drill-and-practice) or multiple forms of learning (e.g. collaborative, exploratory)? | • Is your system's pedagogy appropriate for the target group (e.g. does it detect, adapt to, and build on students' strengths rather than speaking to their weaknesses?)? | • Is your system addressing a gap in AIED support provision or fitting into/exploiting a trend? |

| | DIMENSIONS OF AIED DESIGN AND DEPLOYMENT | | | |
| --- | --- | --- | --- | --- |
| | **AIED Data** | **AIED Models/Algorithms** | **AIED Pedagogy** | **Socio-technical context** |
| | neuro-divergent students may perform better in an exploratory or collaborative mode than in drill and practice)? | • Are your models open to exploration and calibration by users with respect learning modes? | • Are your system pedagogy open to calibration and authoring by practitioners? | |
| | Consequences/Risks: | | | |
| | • Harm of representation<br>• Harm of allocation<br>• Outcome fairness | • Harm of representation<br>• Harm of allocation<br>• Outcome fairness<br>• Process fairness | • Harm of representation<br>• Harm of allocation<br>• Outcome fairness<br>• Human autonomy | • Outcome fairness<br>• Process fairness |

## 6. Conclusions

AI has been and continues to be presented as an opportunity for successfully tackling global education challenges. Given the COVID-19 pandemic, recent discussions have focused on how AI might progress education by enabling teaching and learning online. Companies have scrambled to gain entry into a lucrative educational technology market, spurring calls and demand for EdTech as the foundational infrastructure for accessible remote learning (Williamson, 2021). Popular media and online/offline spaces representing AIED stakeholders (individuals, groups, and organisations, for example) suggest general acceptance that AI *is* the future of education and AI *will* provide a solution for achieving the Sustainable Development Goal 4: "Ensure inclusive and equitable quality education and promote lifelong learning opportunities for all."

However, throughout this chapter, we have discussed diverse questions and challenges related to the ethics of AI and AIED more specifically. While AI can certainly be of benefit to education, there are social, environmental, and ethical concerns that cannot be ignored when weighing the value of AI *for* education, especially when the implementation of AI can result in unfair practices that centre on who has access to these new technologies, or who has access to education at all. These kinds of ethical challenges, if left unexamined by the AIED community, can potentially result in perpetuating and maintaining forms of bias against historically oppressed bodies (Eubanks, 2018; Noble, 2018; Reuters, 2018; O'Neil, 2017; ProPublica, 2016).

In this chapter we have focused only on a subset of ethical considerations of relevance to AIED. There are further ethical challenges related to data management; liability; data ownership; data privacy, and to who sets the educational agenda. For example, related to liability issues, if students' personal data were sold to a third party without the knowledge of the student, it is not clear who should be accountable: Is it the system's commercial developer? Is it the teacher? Or is it the school? The pandemic has also rendered in stark relief that educational technologies for remote learning have been largely concentrated in a handful of transnational technology companies such as Microsoft, Facebook, Boxlight, and Amazon. Two primary concerns arise here: (i) the concentration of personal (student and teacher) information might create a privacy risk (e.g. attracting malicious parties); and (ii) these dominant platforms could form a monopoly to dominate the market on research and development of algorithms (UNESCO, 2019), ultimately having the power to dictate, or at least strongly influence, educational policies and practices. Related to this are the distributed service architectures used as foundational for algorithms being integrated and deployed across education. These architectures comprise feedback in real-time from users and third-party service providers, which is used for a range of optimisation activities designed to extract value through the system. While optimisation tends to be used for technical performance and reduction of costs, it has also become a part of the allocation of continuous development to build adaptive systems. The issue is that such optimisations are designed to capture detailed information about individuals and groups and to manipulate behaviours. In this capacity, such systems introduce to education risks and harms that extend beyond bias and discrimination. As also discussed by Williamson et al. (*in this volume*), focusing only on algorithms misses the results of optimising all aspects of a broader socio-technical system, with discrimination becoming one of the injustices that emerge when these systems are designed and developed to maximise profits. Failing to attend to incentives of service providers and their ability to enforce solutions constrains our understanding of how they operate and fixes in certain actors the power to shape decisions and behaviours which have significant impacts on society at large (Kulynych et al, 2018).

Many of the issues related to optimisation are beyond the scope of this chapter. Nevertheless, it is critical for the AIED community to be cognisant of the context in which its technologies are being developed, deployed, and exploited. Trying to understand this context is critical to the community and individual designers being able to appraise the role of their work in the broader socio-technological context and to being able to question the ethical value of the methods they use and the systems they create beyond the declared good intentions. A comprehensive understanding of the ethics of AIED needs to involve horizon scanning and interdisciplinary conversations, explicitly taking into account insights from the learning sciences, cognitive and educational neuroscience, the sociology of education, and philosophical introspection. It also calls for the community to consider fundamental questions about what education is for, what kind of educational systems does AIED as field support (and whether that in itself is ethical), and how the community might be able to shape, through engaging in ethics by design, the beneficence of education for all. All of these are necessary to help us identify and explore the *unknown unknowns* of the ethics of AIED, in order to establish a balanced, self-aware, and above all – actionable approach to *Ethical AIED* practices that are able to bridge between the ethical implications of AIED systems' design decisions at the algorithmic micro level and the key considerations at the socio-technological macro level.